\documentclass[10pt]{paper}

\usepackage[T1]{fontenc}
\usepackage{textcomp}
\usepackage{yfonts}

\usepackage{color}

\IfFileExists{graphicx.sty}{\usepackage{graphicx}}{}
\usepackage{amsmath}
\usepackage{amsthm}
\usepackage{amsfonts}
\usepackage{amssymb}
\usepackage{mathpazo}
\usepackage{stmaryrd}
\usepackage{hyperref}
\usepackage{boxedminipage}
\usepackage{paralist}
\usepackage{multirow}
\usepackage{color}
\usepackage{cancel}
\usepackage{pdfpages}
\usepackage{synttree}
\usepackage{enumitem}
\setlist[itemize]{nosep,label=--}
\usepackage{tocloft}

\usepackage[margin=2cm]{geometry}

\definecolor{RedLetter}{rgb}{0.63,0.165,0.163}
\definecolor{dim}{rgb}{0.2,0.2,0.2}
\definecolor{pink}{rgb}{1.0,0.8,0.8}

\DeclareMathSymbol{\N}{\mathbin}{AMSb}{"4E}
\DeclareMathSymbol{\Z}{\mathbin}{AMSb}{"5A}
\DeclareMathSymbol{\R}{\mathbin}{AMSb}{"52}
\DeclareMathSymbol{\Q}{\mathbin}{AMSb}{"51}
\DeclareMathSymbol{\I}{\mathbin}{AMSb}{"49}

\usepackage{mathrsfs}

\theoremstyle{definition}

\newcommand{\editor}[1]{\begin{sc}\color{RedLetter} #1\end{sc}}
\newcommand{\ignore}[1]{}

\newcommand{\reffig}[1]{Figure~\ref{#1}}
\newcommand{\refsec}[1]{Section~\ref{#1}}

\usepackage[all,color]{xy}

\usepackage{float}
\floatstyle{boxed}
\restylefloat{figure}

\definecolor{RedLetter}{rgb}{0.63,0.165,0.163}

\newcommand{\mytitle}{Incremental Maintenance for Leapfrog Triejoin}

\title{\mytitle}
\author{Todd Veldhuizen\thanks{LogicBlox Inc., \texttt{tveldhui@acm.org}}}

\date{\today}

\begin{document}

\maketitle

\begin{abstract}
We present an incremental maintenance algorithm for 
leapfrog triejoin.  The algorithm maintains rules in
time proportional (modulo log factors) to the edit distance
between leapfrog triejoin traces.
\end{abstract}

\setcounter{tocdepth}{2}
\tableofcontents

\ignore{***
\section{To Do}
\begin{itemize}
\item Formalize complexity argument
\item Handling of fixpoints?
\item Benchmark results, e.g., 4-clique
\item \editor{There must be DB papers on incrementally maintaining scans?}
\item Periodic reoptimization of rules
\end{itemize}
***}

\newpage

\section{Introduction}

Incremental evaluation is a perennial topic in computer science.
The basic problem is easily described: given an expensive computation,
and some change to its inputs, we want to efficiently update the
result, without recomputing from scratch.

The most common traditional approach to such problems is to consider a
computation graph, in which edges carry values, and vertices
represent small computations, inputs or outputs.
For example, the arithmetic expression
$r=(a+b)*(c+d)$ could be represented by this graph:
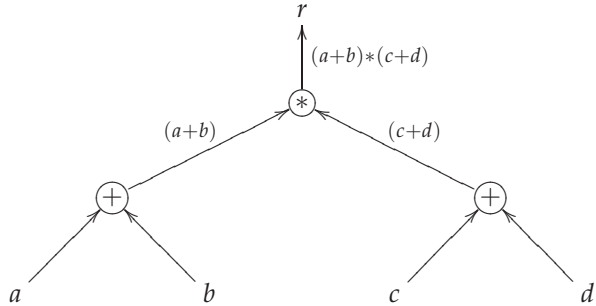
\newcommand{\xynode}[1]{*+[o][F-]{#1}}
\begin{align*}
\xymatrix {
& & & r & \\
& & & \xynode{\ast} \ar[u]_{(a+b)*(c+d)} \\
& \xynode{+} \ar[urr]^{(a+b)} & & & & \xynode{+} \ar[ull]_{(c+d)} & & & &\\
a \ar[ur] & & b \ar[ul] & & c \ar[ur] & & d \ar[ul]
}
\end{align*}
We can regard the above graph as a \emph{trace}, i.e.,
a low-level history of the computation.
When an input such as $b$ changed, the effect of the change
can be rippled through
the graph, only re-evaluating those vertices whose inputs change.
We can call this process \emph{trace maintenance}, and if done
properly, can be done in time cost proportional to the
number of changes required to update the trace.

In databases the problem of incremental evaluation is known as
\emph{incremental view maintenance}.
A view is simply a query installed in a database,
and kept up-to-date as its input relations change.

Incremental view maintenance has historically been done using
one of two techniques \cite{Chen:2005}:
\begin{enumerate}
\item Syntactic approaches derive special rules 
to update a view.  For example, given a rule
such as:
\begin{align*}
C(x) &\longleftarrow A(x), B(x).
\end{align*}
which computes the intersection $A \cap B$,
one can automatically derive rules to update the view when
elements are inserted to A or B.  For example, a rule which
says `If x is inserted to $A$, and $x \in B$, then insert
$x$ into $C$' can be written:
\begin{align*}
+C(x) &\longleftarrow +A(x), B(x)
\end{align*}
There are two challenges associated with this approach:
(a) for complex rules one can encounter a combinatorial
explosion of update rules; and (b) the update rules may be
difficult to evaluate efficiently.  In particular, any
claim of efficiency for this approach must resort to
a \emph{deus ex machina} appeal to the strength of the
query optimizer.
\item Algebraic approaches follow a trace-maintenance
approach, at a coarse level of granularity, where each vertex
in the graph represents an algebra operator (e.g., join, projection).
One defines special maintenance algorithms for each operator,
so that e.g. a projection can be maintained efficiently when its
input relation is updated.
\end{enumerate}

In this paper we present an incremental maintenance
algorithm for leapfrog triejoin \cite{Veldhuizen:LB:2012}, a join algorithm
with worst-case optimality guarantees.
This maintenance algorithm is implemented in the
Delve runtime engine of our
commercial Datalog system LogicBlox\textsuperscript{\textregistered}.

Our approach to incremental maintenance is rather different
from the usual database approaches, and is loosely inspired by
the dynamization procedure of Acar et al \cite{Acar:SODA:2004}.
It hews to the \emph{trace maintenance}
approach, arguably the traditional technique in computer science.
Unlike the algebraic approaches mentioned above,
we maintain the trace at a very fine level of granularity---at the
level of individual iterator operations
in the leapfrog triejoin algorithm.
Our maintenance algorithm has time cost
proportional to trace distance (modulo log factors),
giving it an optimality guarantee.

\subsection{Aspiration}

Suppose we have some Datalog rule, for example:
\begin{align*}
F(x,y) &\longleftarrow G(x,z),H(y,z),I(x,y,z).
\end{align*}
After evaluating this rule to calculate $F(x,y)$, some transaction(s) are committed
that modify $G$, $H$, and/or $I$, and we wish to update $F$ to reflect
these changes.  In many situations it is prohibitively expensive to
recalculate $F$ from scratch, so we instead aim to efficiently \emph{maintain}
$F(x,y)$ based on the changes made to the predicates.

The aspiration of our maintenance algorithm is
\emph{maintenance cost proportional to trace edit distance}.
Unpacking this a bit:
\begin{itemize}
\item By maintenance cost, we mean the number of steps required
to maintain a rule, i.e. produce new versions of the head predicates
in response to changes made to body predicates.
\item By trace, we mean a low-level step-by-step description of
the operations performed during full evaluation of a rule,
i.e., a succinct history of the computation.
We maintain traces at the level of predicate iterator operations,
so these steps might include items such as
``position the iterator
for $G(x,z)$ at a least upper bound for $(x=1531,z=142)$''.
\item By \emph{trace edit distance}, we mean comparing side-by-side
the trace for full-evaluation on the original predicates (e.g. $G,H,I$)
with the trace for full-evaluation on the modified predicates
(e.g. $G',H',I'$), and counting how many changes must be made
to the original trace to turn it into the trace for full-evaluation
on the modified predicates.
\end{itemize}

For a trivial illustration, suppose I have a rule
\begin{align*}
C[x]\mbox{=}z &\longleftarrow z\mbox{=}A[x]\mbox{+}B[x].
\end{align*}

If I evaluate this rule in a hypothetical debugging mode where each
step is logged, I might get a table like this:
\begin{quote}
\begin{verbatim}
   x  | A[x]  B[x]  C[x]
   0  |   0     0     0
   1  |  10     0    10
   2  |   0     1     1
   3  |  30     1    31
   4  |   0     0     0
   5  |   0     0     0
\end{verbatim}
\end{quote}
This table is a rough approximation of what we mean by a `trace': a
step-by-step description of the evaluation.
Now suppose I make some changes to $A[x]$ and do another full evaluation,
which produces this table (with differences marked by an asterisk):
\begin{quote}
\begin{verbatim}
   x  | A[x]  B[x]  C[x]
   0  |   0     0     0
   1  |  10     0    10
   2  |  20*    1    21*
   3  |  30     1    31
   4  |   0     0     0
   5  |  50*    0    50*
\end{verbatim}
\end{quote}
If I run the unix command 'diff' on these two tables, I get:
\begin{quote}
\begin{verbatim}
   < 2  |   0     1     1
   > 2  |  20     1    21
   < 5  |   0     0     0
   > 5  |  50     0    50
\end{verbatim}
\end{quote}
The length of this diff hints at what we mean by `trace edit distance':
the number of changes (edits) you'd need to make to the original trace
to turn it into the trace of full-evaluation on the modified predicates.

There is a large history to the general problem of
incremental maintenance (not just for Datalog) that says
this goal is achievable; the challenge is
finding a solution that achieves this goal yet \emph{performs well}.

\subsection{Summary of the maintenance algorithm}

We give a brief sketch of our approach, for orientation.

First, some notations.
Given two versions $C,C'$ of a predicate $C(x)$, we write $(C \cdots C')$
for the \emph{difference} between the two versions.  
We mean by this a `delta' relation of the form:
\begin{align*}
(C \cdots C')(x,\Delta) &= 
\begin{cases}
(x,\mbox{INSERT}) & \text{if } x \not\in C \text{ and } x \in C' \\
(x,\mbox{ERASE})  & \text{if } x \in C \text{ and } x \not\in C'
\end{cases}
\end{align*}
i.e. a predicate enumerating the differences between $C,C'$,
with the $\Delta$ variable taking on values $\mbox{INSERT}$
and $\mbox{ERASE}$.

Consider the rule mentioned above:
\begin{align*}
F(x,y) &\longleftarrow G(x,z),H(y,z),I(x,y,z).
\end{align*}

Let $\mathsf{Body}[G,H,I](x,y,z) = G(x,z),H(y,z),I(x,y,z)$
be the body of the rule.  The basic approach to maintenance
is to evaluate a rule of the form:
\begin{align*}
\delta F(x,y,\Delta) &\longleftarrow 
\begin{array}[t]{l}
(\mathsf{Body}[G,H,I] \cdots \mathsf{Body}[G',H',I'])(x,y,z,\Delta), \\
\mathsf{ChangeOracle}(x,y,z).
\end{array}
\end{align*}

\begin{itemize}
\item The left-hand side $\delta F(x,y,\Delta)$ gives a set of changes
to be applied to the predicate $F$ to produce the updated predicate
$F'$.
\item The term
$(\mathsf{Body}[G,H,I] \cdots \mathsf{Body}[G',H',I'])(x,y,z,\Delta)$
enumerates differences in the satisfying assignments of the
rule-body for $G,H,I$ vs. $G',H',I'$.  This is done by simply
evaluating both bodies and comparing the results.
\item The
$\mathsf{ChangeOracle}(x,y,z)$ term serves to restrict
evaluation to just those regions of the tuple-space where
changes might occur.  The maintenance rule would be
correct (but inefficient) if $\mathsf{ChangeOracle}(x,y,z)$
were omitted.
\end{itemize}

There are two primary tasks:
\begin{enumerate}
\item How to represent head predicates, and how to update them given deltas.
We describe these techniques in \refsec{s:heads}, and some specialized
data structures and algorithms in \refsec{s:algorithms}.
\item How to construct and employ the $\mathsf{ChangeOracle}$ predicate.
This is described in \refsec{s:oracle}.
\end{enumerate}

\subsection{Background, terminology, and notations}

Our variant of Datalog supports both relations such as
$R(x_1,\ldots,x_k)$ and functions such as $F[x_1,\ldots,x_k]=y$.
We refer to functions and relations as \emph{predicates}.
A relation or function symbol together with its arguments is
an \emph{atom}.

\begin{figure}
\begin{align*}
\mathsf{conj} &::= ~[~ \exists \overline{x} ~.~ ~]~ \mathsf{dform} ~[~, \mathsf{dform} \cdots ~]~ \\
\mathsf{dform} &::= \mathsf{atom} ~|~ \mathsf{disj} ~|~ \mathsf{negation} \\
\mathsf{atom} &::= R(\overline{y}) ~|~ F[\overline{y}]=\overline{z} \\
\mathsf{disj} &::= \mathsf{conj} ; \mathsf{conj} [ ; \mathsf{conj} \cdots ] \\
\mathsf{negation} &::= ~! \mathsf{conj} \\
\\
\mathsf{rule} &::= \forall \overline{x} ~.~ \mathsf{head} \leftarrow \mathsf{conj} \\
\mathsf{head} &::= \mathsf{atom} ~[~ , \mathsf{atom} \cdots ] 
\end{align*}
\caption{\label{f:grammar}Internal representation of rules.}
\end{figure}

A Datalog rule is written in the form $\mathit{head} \leftarrow \mathit{body}.$
A rule head contains one or more atoms; a body is a first-order formula.
Users write rules in a relaxed form without quantifiers, for example:
\begin{align*}
S(x,y) &\leftarrow A(x,y),B(y,z).
\end{align*}
Internally, rules are represented in the more restricted form of
\reffig{f:grammar}, with explicit quantifiers; for example:
\begin{align*}
\forall x, y ~.~ S(x,y) \leftarrow \exists z ~.~ A(x,y),B(y,z) 
\end{align*}
Variables occurring in both head and body are placed in the
rule-level universal quantifier block (e.g. $x,y$); variables
occurring only in the body are ascribed to the existential
quantifier block of the smallest conjunction ($\mathsf{conj}$)
encompassing their uses (e.g. $z$).

A \emph{materialized predicate} is one whose elements are stored
in a data structure.
Materialized predicates can be either \emph{extensional} (EBD) or
\emph{intensional} (IDB):
\begin{itemize}
\item An extensional (EDB) predicate is one whose contents can be
directly manipulated by transactions that insert and remove records.
\item An intensional (IDB) predicate (aka \emph{view}) is defined 
by one or more Datalog rules.  IDB predicates are maintained
incrementally in response to changes made to EDB predicates.
\end{itemize}

A \emph{primitive} is a function or relation that is calculated
on demand.  For example, the function $\mbox{\sf add}[x,y]=z$
is a primitive that computes $z=x+y$.

\subsubsection{Key- and value-position}
For a materialized predicate atom $R(x_1,\ldots,x_k)$ or $F[x_1,\ldots,x_k]=y$,
we say the variables
$x_1,\ldots,x_k$ appear in \emph{key-position}.  (If $F$ is a
primitive operation, e.g., $\mathsf{add}[x_1,x_2]=y$, we do not count it
as having key-position appearances of variables.)

A variable is a deemed a \emph{key} if it appears anywhere in key-position
in the body of a rule; otherwise it is a \emph{value}.
A binding for the key variables of a rule uniquely determines the values.
For example, in the expression $F[x]=a,G[y]=b,r=a+b$,
the variables $x,y$ are keys, and the variables $a,b,r$ are values.

\section{Maintaining head predicates}

\label{s:heads}

In this section we describe how to maintain head predicates as
changes are made to satisfying assignments of the body.

\subsection{Projection-free rules}

A rule is \emph{projection-free} if each atom in the head
contains an appearance of every key variable.  For example:
\begin{align*}
\forall x,y,z ~.~ R(x,y,z) &\leftarrow A(x,y),B(y,z)
\end{align*}
is projection-free, whereas:
\begin{align*}
\forall x,y ~.~ S(x,y) &\leftarrow \exists z ~.~ A(x,y),B(y,z)
\end{align*}
is not, because the key-variable $z$ does not appear in the
head atom $S(x,y)$.

Maintaining head predicates for projection-free rules is
easy: we simply insert or remove records in response to the
changes made to satisfying assignments of the body.

\subsection{Rules with projection}

For rules with projection we primarily use counting \cite{Gupta:SIGMOD:1993}.
Consider the rule:
\begin{align*}
S(x,y) &\leftarrow A(x,y),B(y,z)
\end{align*}

We represent $S$ by a predicate $S[x,y]=\eta$, where $\eta$ is a
\emph{support count}:
the number of satisfying assignments of the body producing $(x,y)$.
(In the example rule, there might be several bindings of $z$ for
a given $(x,y)$.)
Then, for $\delta S(x,y,\Delta)$, we respond to a $\Delta=\mbox{INSERT}$
by incrementing $\eta$, and to a $\Delta=\mbox{ERASE}$ by decrementing
$\eta$.  We use special data structure support (an \emph{update-action})
that treats $\eta$ as a
reference count, so that a decrement of $\eta$ resulting in 
$\eta=0$ causes the record to be deleted.

Functions appearing in rule heads are handled in a similar way:
suppose the head predicate is $F[s]\mbox{=}t$.  We maintain the head predicate
as $F[s]\mbox{=}(t,\eta)$, where $\eta$ is the support count.
Given a set of deltas to apply, we order
them so ERASE actions are applied first, to avoid issues with
conflicting function values.

\subsubsection{Short-circuit evaluation}

In some cases we can avoid the use of reference counts by using
\emph{short-circuit evaluation}.  For a rule such as:
\begin{align*}
S(x,y) &\leftarrow A(x,y),B(y,z)
\end{align*}
it is helpful to explicitly insert a quantifier for $z$:
\begin{align*}
S(x,y) &\leftarrow \exists z . A(x,y),B(y,z)
\end{align*}
Suppose the key order chosen by the optimizer is $[x,y,z]$.
Given particular $x,y$, it is obviously of little use to enumerate 
all possible satisfying assignments for $z$.  We can instead use
short-circuit evaluation: as soon 
the first satisfying assignment for an $(x,y)$ is
produced, we can backtrack immediately without considering further
assignments of $z$.  In this case, the support count $\eta$ is
unnecessary.  However, in some cases it might be more efficient
to use a key order such as $[z,y,x]$, in which case short-circuit
evaluation cannot be used.  This decision is left to the query optimizer.

\subsection{Aggregations}

Our variant of Datalog supports aggregations such as
sum, count, min, and max.  For example, the following
rule computes the total calories consumed by people
from meals:
\begin{align*}
\begin{array}{l}
\mathsf{CaloriesConsumed}[\mathit{person}]=\mathit{totcal} \leftarrow \\
~~~\mathsf{agg} \ll \mathit{totcal}=\mathsf{sum}(\mathit{cal}) \gg \\
~~~~~~\mathsf{ate}(\mathit{person},\mathit{meal}),\mathsf{caloriesOf}[\mathit{meal}]=\mathit{cal}.
\end{array}
\end{align*}

\subsubsection{Aggregations: count}

Count aggregations can be handled by using the same mechanism used
for support counts of rules with projections.  For example:
\begin{align*}
\begin{array}{l}
\mathsf{outdegree}[x]=d \leftarrow \\
~~~\mathsf{agg} \ll d=count() \gg \\
~~~~~~E(x,y).
\end{array}
\end{align*}
can be implemented using a head predicate $\mathsf{outdegree}[x]=d$,
where $d$ is a support count as described above.
The support count is incremented and decremented in response to
changes in satisfying assignments of the rule body:
if a new satisfying assignment $E(x,y)$ is found, then
$d$ is incremented; if a satisfying assignment is removed,
then $d$ is decremented, and the record is removed from
$\mathsf{outdegree}$ when $d=0$.

\subsubsection{Aggregations: the Abelian group case}

\label{s:Abelian}

For sum aggregations over an Abelian group $\langle G,+,-,0\rangle$,
where the operator $+$ associative and commutative,
we can use an update-action that updates the aggregate by
'adding' the new value when $\Delta=\mbox{INSERT}$, and
'adding' the inverse ('negative') of the new value when
$\Delta=\mbox{ERASE}$.  We also employ a support count $\eta$
to remove a record once no more satisfying assignments of
the body contribute to it.

This style of aggregation can be used for sum aggregations
over integers and fixed-precision data types.

\subsubsection{Aggregations: the semigroup case}

\label{s:semigroup}

Min and max aggregations cannot be treated as Abelian group aggregations,
since there is no inverse:  i.e. no operation
$\cdot{}^{-1}$ such that $\forall \alpha ~.~ \min(\alpha^{-1},\alpha)=I$,
where $I=-\infty$ is an identity element.
We instead treat these as aggregations
over a semigroup $(M,\oplus)$, where $\oplus$ is a binary
operator (e.g., min, max).

Consider the example aggregation:
\begin{align*}
\begin{array}{l}
A[x]=ms \leftarrow \\
~~~\mathsf{agg} \ll ms=max(s) \gg \\
~~~~~~D[x,y,z]=s.
\end{array}
\end{align*}

We use an intermediate predicate that
supports scans, as described in \refsec{s:scans}.  Each
satisfying assignment of the body is inserted into this
intermediate predicate by a special rule:
\begin{align*}
\begin{array}{l}
A_{\mathit{max-scan}}[x,y,z]=s \leftarrow\\
~~~ D[x,y,z]=s.
\end{array}
\end{align*}
The head predicate $A[x]=ms$ is computed by performing scans
on the $A_{\mathit{max-scan}}$ predicate, which
we can write:
\begin{align*}
\begin{array}{l}
A[x]=\mathbf{Scan}\left( A_{\mathit{max-scan}}, [x,-\infty,-\infty], [x,+\infty,+\infty] \right)
\end{array}
\end{align*}
That is, for each $x$, $A[x]$ is computed by taking a scan of all
records in the interval from $[x,-\infty,-\infty]$ to $[x,+\infty,+\infty]$,
where $-\infty, +\infty$ are the smallest/largest representable values of
the datatype.

For each change in satisfying assignments of the body,
we insert or remove records to/from the intermediate predicate
$A_\mathit{max-scan}[x,y,z]=s$, and then recompute whatever records of
$A[x]=ms$ could have changed.
This lets us maintain the aggregation result in time
$O(\delta \log n)$, where $\delta$ is
the number of changes in satisfying assignments of the body.

Note: we can reuse the intermediate predicate
$A_{\mathit{max-scan}}[x,y,z]=s$ to provide aggregations
at multiple levels of detail.
For example, if we also wanted to know the maximum $s$ for a
given $x,y$ pair, we could define a rule:
\begin{align*}
\begin{array}{l}
A'[x,y]=ms \leftarrow \\
~~~\mathsf{agg} \ll ms=max(s) \gg \\
~~~~~~D[x,y,z]=s.
\end{array}
\end{align*}
which could share the $A_{\mathit{max-scan}}[x,y,z]=s$
intermediate predicate with the rule calculating $A[x]=ms$:
\begin{align*}
\begin{array}{l}
A'[x,y]=\mathbf{Scan}\left( A_{\mathit{max-scan}}, [x,y,-\infty], [x,y,+\infty] \right)
\end{array}
\end{align*}

String concatenation aggregations can also be handled using
the semigroup approach; this can be made efficient by representing
long strings using ropes \cite{Boehm:SPE:1995}.

\subsubsection{Aggregations: floating-point sums}

Floating-point sum aggregations are problematic because floating-point
addition is not associative.  The Abelian group approach described above
would allow arbitrarily large errors to accumulate over time as the sum
was maintained.  The
semigroup approach would produce answers that depended in subtle ways
on the precise structure of the scan-tree (\refsec{s:scans}), due to
nonassociativity; it would also have the undesirable requirement of
storing all satisfying assignments of the rule body in an intermediate
data structure.

The sensible alternative is to employ a head-predicate with an arbitrary-precision
floating-point value, which lets us use the Abelian group approach.
That is, for an aggregation such as:
\begin{align*}
\begin{array}{l}
\mathsf{F}[x]=\mathit{tot} \leftarrow \\
~~~\mathsf{agg}\ll \mathit{tot}=total(v) \gg \\
~~~~~~G[x,y]=v
\end{array}
\end{align*}
where $v$ is a floating-point value, we use an intermediate predicate of the
form $F^\ast[x]=(\mathit{tot}^\ast,\eta)$, where $\mathit{tot}^\ast$ is represented
using an arbitrary-precision type, and use the update technique mentioned
in \refsec{s:Abelian}.

There is a useful trick that can be employed here to efficiently represent
$\mathit{tot}^\ast$.  Consider a floating-point sum $S=\sum_{i \in I} s_i$.
Instead of representing $S$ directly, we can instead represent
the sum $X+\left(\sum_{i \in I} s_i\right)$, where $X \in \R$ is a value
such as:
\begin{align*}
X &= \sum_{k=-512}^{512} 2^{4k}
\end{align*}
The binary representation of $X$ is e.g.:
\begin{align*}
10001000100010001000\cdots1000100010001.000100010001000\cdots1000100010001
\end{align*}
We partition $X$ into 52-bit segments, this being the number of mantissa bits
in an IEEE 754 double-precision floating point number.  We only store
52-bit segments of $X + \left(\sum_{i \in I} s_i\right)$ that differ from the
corresponding segment of $X$, representing each segment as a floating-point
number.
Since $X$ has $1$-bits at regular intervals, any borrowing required to
accommodate a negative summand never requires increasing the Hamming distance
between $X$ and $X + \left(\sum_{i \in I} s_i\right)$ by more than 4 bits.
(Consider for example representing the sum of
$S=\{ 2^{500}, -1 \}$: with this representation we do not have to borrow
from $2^{500}$, which would cause a run of 500 1's in the representation;
instead we just borrow from $2^4$.  We would store only two segments, the
one containing $2^{500}$ and the one containing $2^0$.)

To extract $F[x]=\mathit{tot}$ from the intermediate predicate
$F^\ast[x]=(\mathit{tot}^\ast,\eta)$, we use a rule of the form:
\begin{align*}
\begin{array}{l}
F[x]=\mathit{tot} \leftarrow\\
~~~F^\ast[x]=(\mathit{tot}^\ast,\eta),\\
~~~\mathit{tot}=\mathsf{toFloat}[\mathit{tot}^\ast].
\end{array}
\end{align*}
The primitive $\mathsf{toFloat}$ is straightforward to implement:
we identify the first bit-position where 
$X + \left(\sum_{i \in I} s_i\right)$ differs
from $X$.  The value $\mathit{tot}$ is positive if the first differing bit is zero,
and negative if the first differing bit is one.
We subtract $X$, and extract a 52-bit mantissa.  Combined with an exponent
and sign, this yields a double-precision floating-point quantity.

\section{Algorithms \& Data structures}

\label{s:algorithms}

\subsection{Scans}

\label{s:scans}

Scans (prefix-sums) are a handy formalism for
aggregation-like operations \cite{Chatterjee:SUPER:1990}.  We
employ them for semigroup aggregations (\refsec{s:semigroup}),
and also for implementing queries on sensitivity indices
(\refsec{s:sensitivityindices}).

Given an array
$A=[a_1,a_2,\ldots,a_n]$ and an associative
operator $\oplus$, the scan of $A$ is just
$a_1 \oplus a_2 \oplus \cdots \oplus a_n$.
If we choose $\oplus$ to be addition, we get
the sum; if we choose $\oplus$ to be the
$\max$ operator, we get the maximum element.

Suppose we want to calculate the aggregation over an arbitrary
interval, i.e.  $a_i \oplus \cdots \oplus a_j$
where $1 \leq i \leq j \leq n$.  (For example, if
the $A$ array contained sales values for each day,
we might want to aggregate sales over a specific
month, rather than over all time.)

Associativity of the $\oplus$ operator permits
a simple data structure that can calculate
the scan of any interval in $O(\log n)$ time.
Let $A_{ij} = a_i \oplus \cdots \oplus a_j$
be the aggregation over the elements $a_i,\ldots,a_j$.
We construct a binary tree (a \emph{scan-tree})
with each leaf a
single element $a_i$, and each internal node
storing the $\oplus$-sum of its children:
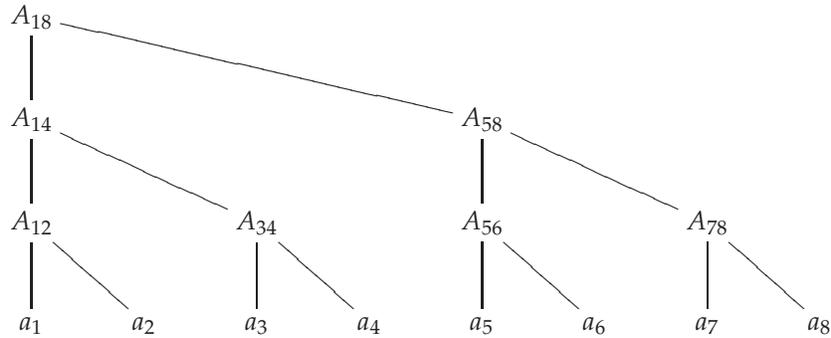
\begin{align*}
\xymatrix{
A_{18} \ar@{-}[d] \ar@{-}[drrrr] \\
A_{14} \ar@{-}[d] \ar@{-}[drr] & & & & A_{58} \ar@{-}[d] \ar@{-}[drr] \\
A_{12} \ar@{-}[d] \ar@{-}[dr] & & A_{34} \ar@{-}[d] \ar@{-}[dr] & & A_{56} \ar@{-}[d] \ar@{-}[dr]& & A_{78} \ar@{-}[d] \ar@{-}[dr] \\
a_1 & a_2 & a_3 & a_4 & a_5 & a_6 & a_7 & a_8
}
\end{align*}
For example, the left child of the root is:
\begin{align*}
A_{14} &= A_{12} \oplus A_{34} \\
              &= (a_1 \oplus a_2) \oplus (a_3 \oplus a_4)
\end{align*}

To calculate the aggregation of an arbitrary interval 
we take the $\oplus$-sum of all subtrees contained entirely in the interval.
For example:
\begin{align*}
a_1\oplus\cdots\oplus a_8 &= A_{18} \\
a_1\oplus\cdots\oplus a_3 &= A_{12} \oplus a_3 \\
a_2\oplus\cdots\oplus a_8 &= a_2 \oplus A_{34} \oplus A_{58} \\
a_3 \oplus \cdots \oplus a_7 &= A_{34} \oplus A_{56} \oplus a_7 \\
a_4 \oplus \cdots \oplus a_7 &= a_4 \oplus A_{56} \oplus a_7
\end{align*}
For any interval $a_i,\ldots,a_j$, we never need to sum more than
$2 \lceil \log_2 n \rceil = O(\log n)$ elements.

If a value changes, say $a_5$ is changed to $a_5'$,
we can update the scan-tree by simply recalculating all internal
nodes on the path from $a_5$ to the root ($A_{56}, A_{58}, A_{18}$).
This requires only $O(\log n)$ operations.

\newcommand{\dlagg}[2]{
   \begin{array}[t]{l}
   \mathbf{agg}\langle \hspace{-0.3em} \langle #1 \rangle \hspace{-0.3em} \rangle \\
   \hspace{0.25in}
      \begin{array}[t]{l}
      #2
      \end{array}
   \end{array}
}
\newcommand{\dlvar}[1]{\mathit{#1}}
\newcommand{\dlpred}[1]{\mathsf{#1}}

For a concrete example, suppose we have a predicate
$\dlpred{sales}[\dlvar{region},\dlvar{store}]\mbox{=}\dlvar{tot}$ giving 
the total sales for each store, and we wish to maintain the maximum
sales of any store in each region:
\begin{align*}
\dlpred{maxsales}[\dlvar{region}]\mbox{=}\dlvar{maxtot} &\longleftarrow
   \dlagg{\dlvar{maxtot}\mbox{=}\max(\dlvar{tot})}{\mathsf{sales}[\dlvar{region},\dlvar{store}]\mbox{=}\dlvar{tot}.}
\end{align*}

Shown below is a scan-tree for calculating the $\max$-aggregation
of a $\dlpred{sales}$ predicate with 16 records.  The first three
columns give the (region,store,tot) records, and the scan-tree is
drawn to the right.
The records relevant for region 2 have been highlighted:
\begin{align*}
\xymatrix @=0.5pc {
\dlvar{region} & \dlvar{store} & \dlpred{sales}[\dlvar{region},\dlvar{store}] \\
1 & 1 & 1000.00 & \ar@{-}[l] \ar@{-}[ld] 1500.00 & 8000.00 \ar@{-}[l] \ar@{-}[ldd] & 15000.00 \ar@{-}[l] \ar@{-}[ldddd] & 15000.00 \ar@{-}[l] \ar@{-}[ldddddddd] \\
1 & 2 & 1500.00 \\
1 & 3 & 7300.00 & 8000.00 \ar@{-}[l] \ar@{-}[ld] \\
1 & 4 & 8000.00 \\
1 & 5 & 15000.00 & 15000.00 \ar@{-}[l] \ar@{-}[ld] & 15000.00 \ar@{-}[l] \ar@{-}[ldd] \\
2 & 6 & \mathbf{2900.00} \\
2 & 7 & \mathbf{3500.00} & \mathbf{3500.00} \ar@{-}[l] \ar@{-}[ld] \\
2 & 8 & \mathbf{1440.00} \\
2 & 9 & \mathbf{3300.00} & \mathbf{3300.00} \ar@{-}[l] \ar@{-}[ld] & \mathbf{7024.00} \ar@{-}[l] \ar@{-}[ldd] & 9000.00 \ar@{-}[l] \ar@{-}[ldddd] \\
2 & 10 & \mathbf{1245.00} \\
2 & 11 & \mathbf{7024.00} & \mathbf{7024.00} \ar@{-}[l] \ar@{-}[ld] \\
2 & 12 & \mathbf{5510.00} \\
2 & 13 & \mathbf{9000.00} & 9000.00 \ar@{-}[l] \ar@{-}[ld] & 9000.00 \ar@{-}[l] \ar@{-}[ldd] \\
3 & 14 & 325.00 \\
3 & 15 & 4000.00 & 5300.00 \ar@{-}[l] \ar@{-}[ld] \\
3 & 16 & 5300.00
}
\end{align*}

To calculate the maximum sales for region 2, we can just take the max
of all subtrees for region 2: 
\begin{align*}
\dlpred{maxsales}[2]&=\max(2900.00,3500.00,7024.00,9000.00)\\
&=9000.00
\end{align*}

\subsubsection{Adapting scan-trees for paged data structures}

In practice, we use a Btree-like data structure where leaf and index 
pages are augmented with scan information.  Each leaf page is augmented with a
ScanTree data structure that uses approximately 5\% of the available
space.  It maintains a scan-tree for the records stored on the leaf
page.  This scan-tree uses a binary tree structure, but each scan-tree-leaf
might aggregate a dozen or so records.  For example, the sales
data might be represented on a Btree-like leaf page by this scan-tree:
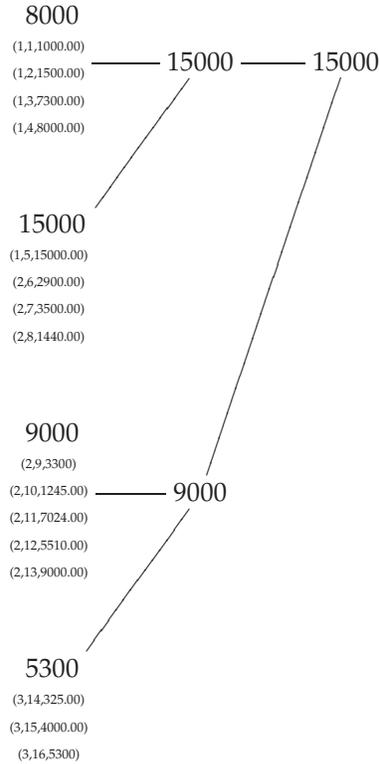
\begin{align*}
\xymatrix{
\txt{ 8000 \\ \tiny (1,1,1000.00) \\ \tiny (1,2,1500.00) \\ \tiny (1,3,7300.00) \\ \tiny (1,4,8000.00)} & 15000 \ar@{-}[l] \ar@{-}[dl] & 15000 \ar@{-}[l] \ar@{-}[ldd] \\
\txt{ 15000 \\ \tiny (1,5,15000.00) \\ \tiny (2,6,2900.00) \\ \tiny (2,7,3500.00) \\ \tiny (2,8,1440.00)} \\
\txt{ 9000 \\ \tiny (2,9,3300) \\ \tiny (2,10,1245.00) \\ \tiny (2,11,7024.00) \\ \tiny (2,12,5510.00) \\ \tiny (2,13,9000.00)} & 9000 \ar@{-}[dl] \ar@{-}[l] \\
\txt{ 5300 \\ \tiny (3,14,325.00) \\ \tiny(3,15,4000.00) \\ \tiny (3,16,5300)}
}
\end{align*}
The records being aggregated (shown in small font) are not actually stored in 
the scan-tree.

Each leaf in the scan-tree aggregates a variable number of records, so that
insertions and deletions can be handled efficiently.  The tree is occasionally
rebalanced to ensure no child aggregates more than twice as many 
records as its sibling.  This permits the scan tree to be updated in
$O(\log B)$ amortized time (where $B$ is the Btree leaf page record capacity)
in response to a record insert/update/delete.

For Btree-style index pages, we augment each record with an extra field
containing scan information.  On a Btree index page, records are typically
of the form $(\mathit{key}_1$,$\ldots$,$\mathit{key}_n$; $\mathit{pageid})$, where
$\mathit{pageid}$ is the page number of a next-level leaf or index page.
We add an additional scan-related field, so records are of the form
$(\mathit{key}_1$,$\ldots$,$\mathit{key}_n$; $\mathit{scan}$, $\mathit{pageid})$,
where $\mathit{scan}$ aggregates all records in the suBtree
reachable at $\mathit{pageid}$.  In addition, each index page
gets a ScanTree (taking approx. 5\% of the page space)
that aggregates the $\mathit{scan}$ elements of the
index-page records.  This approach allows us to calculate the
scan of any interval in an arbitrarily large predicate in
$O(\log n)$ time, where $n$ is the number of records.

\subsubsection{Efficient iteration of complements}

Suppose we have a set $S \subseteq T$, and we wish to iterate the 
complement $T \setminus S$.  This can be done efficiently using
representations for $S$ and $T$ that include a scan-tree for
a 'count' aggregation.  Such a scan tree lets us 
count the number of records in an interval $[k_1,k_2]$ in
$O(\log n)$ time, where $k_1,k_2$ are keys (or key-tuples).

To iterate the complement, we can employ
the principle that if a given key interval
$[k_1,k_2]$ contains the same number of records in $S$ and $T$, then
the complement $T \setminus S$ is empty in that interval.
This reduces the cost of iterating the complement to
$O(|T\setminus S| \cdot \log n)$, a useful improvement for sparse complements.
(The naive approach of iterating $T$ and doing lookups in $S$ would
require $O(|T|)$ time.)

\subsection{Interval trees}

\label{s:intervaltrees}

We use scan trees to implement interval trees,
used to represent \emph{sensitivity indices} in our maintenance
algorithm (\refsec{s:sensitivityindices}).

A simple interval tree stores a set of intervals I,
with each interval of the form $[a,b]$ where $a,b \in K$ and $K$ is some scalar
key type.
An interval query finds the set of intervals
containing some key $x$ of interest, i.e.
\begin{align*}
\mathrm{IntervalQuery}(x) &= \{ [a,b] \in I ~:~ x \in [a,b] \}
\end{align*}

For example, we might have
\begin{align*}
I &= \{ [2,10], [3,7], [5,15], [6,9] \}
\end{align*}
and in response to the query `What intervals contain 10?'
it would produce $\{ [2,10], [5,15] \}$.

To implement an interval tree, we can use a
scan-tree, where each internal node of the
scan tree has a pair $[a,b]$, where $a$ is
the min of the interval starts, and $b$ is
the max of the interval ends.

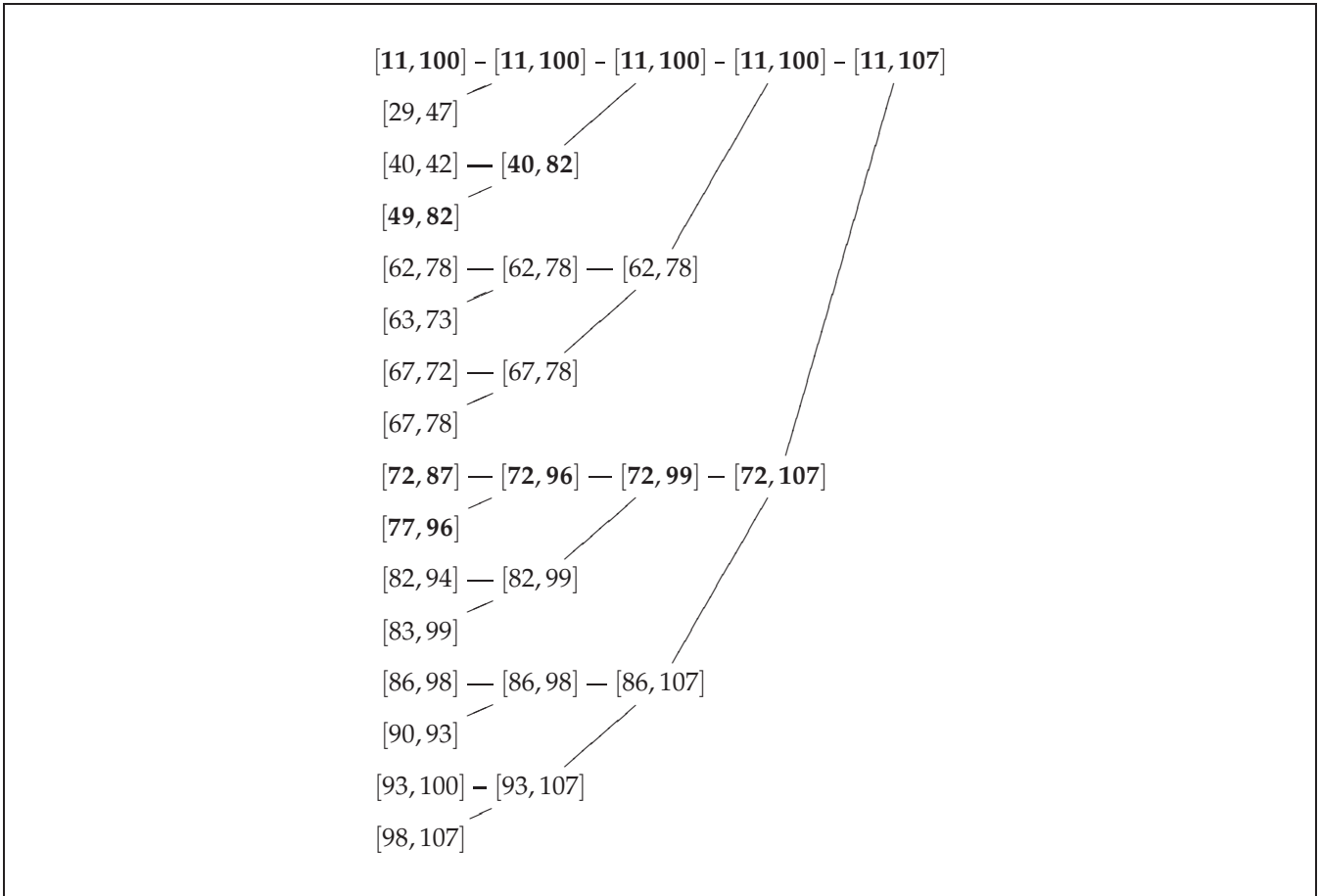
\begin{figure}
\begin{align*}
\xymatrix @=0.3pc {
\mathbf{[11,100]} & \mathbf{[11,100]} \ar@{-}[l] \ar@{-}[dl] & \mathbf{[11,100]} \ar@{-}[l] \ar@{-}[ddl] & \mathbf{[11,100]} \ar@{-}[l] \ar@{-}[ddddl] & \mathbf{[11,107]} \ar@{-}[l] \ar@{-}[ddddddddl] \\
[29,47] \\
[40,42] & \mathbf{[40,82]} \ar@{-}[l] \ar@{-}[dl] \\
\mathbf{[49,82]} \\
[62,78] & [62,78] \ar@{-}[l] \ar@{-}[dl] & [62,78] \ar@{-}[l] \ar@{-}[ddl] \\
[63,73] \\
[67,72] & [67,78] \ar@{-}[l] \ar@{-}[dl] \\
[67,78] \\
\mathbf{[72,87]} & \mathbf{[72,96]} \ar@{-}[l] \ar@{-}[dl] & \mathbf{[72,99]} \ar@{-}[l] \ar@{-}[ddl] & \mathbf{[72,107]} \ar@{-}[l] \ar@{-}[ddddl] \\
\mathbf{[77,96]} \\
[82,94] & [82,99] \ar@{-}[l] \ar@{-}[dl] \\
[83,99] \\
[86,98] & [86,98] \ar@{-}[l] \ar@{-}[dl] & [86,107] \ar@{-}[l] \ar@{-}[ddl] \\
[90,93] \\
[93,100] & [93,107] \ar@{-}[l] \ar@{-}[dl] \\
[98,107]
}
\end{align*}
\caption{\label{f:intervaltree}An interval-tree.  Nodes whose intervals contain $x=80$ are
highlighted.}
\end{figure}

\reffig{f:intervaltree} shows an example. The records are in the first column 
(integer intervals), and the scan-tree to the right.
To find all intervals containing a particular number $x$,
we start at the root and recursively descend to each child,
backtracking when the scan-interval does not contain $x$.
The nodes whose interval contain $x=80$ are highlighted
above.  The result set produced for $x=80$ is
$\{[11,100], [49,82],$ $[72,87], [77,96]\}$.

Interval trees produce the set of containing intervals
for a value $x$ in time $O((m+1) \log n)$, where $m$ is
the number of matching intervals and $n$ is the total number of intervals.

To adapt interval trees for paged data structures, we use
a Btree augmented for scans,
configured for a max-scan on the endpoint of each interval.
Since Btree-type data structures are ordered by key,
and index pages store the least key of their subtrees,
Btrees have a built-in min-aggregation for the
startpoint of each interval.

For general sensitivity indices (\refsec{s:sensitivityindices}),
we use predicates with records of the form 
\begin{align*}
\mathit{SensIndex}(\alpha_1,\alpha_2,\ldots,\alpha_m,a,b,\gamma_1,\ldots,\gamma_k)
\end{align*}
This is understood to represent an interval of \emph{tuples}
beginning at $[\alpha_1,\alpha_2,\ldots,\alpha_m,a]$
and ending at $[\alpha_1,\alpha_2,\ldots,\alpha_m,b]$.
The $\gamma_1,\ldots,\gamma_k$ contain supplemental information described later.

\subsection{Delta-iterators}

\label{s:deltaiterators}

Our current implementation of paged data structures
use copy-on-write page-level versioning.
This allows us to iterate through the
difference between two consecutive versions of a predicate in
$O(\delta \log n)$ time, where $\delta$ is the number of
changes made between the two versions, and $n$ is the
maximum record count of the two versions.
This is done by iterating through
the two versions simultaneously, and skipping any subtrees
common to both versions.\footnote{
A more sophisticated planned data structure,
\emph{cascading trees}, does versioning in a way that
minimizes the number of pages altered.  With cascading
trees, the number of pages that must be examined for
delta-iteration is $O(\delta B^{-1/2} \log \delta)$, where $B$ is
the average leaf-page capacity.  In practice, this means
that e.g. 50 changes, even to widely scattered keys,
will usually be concentrated on a single page.
}

\section{Maintaining rule bodies}

\label{s:oracle}

We now describe our maintenance algorithm for rule bodies.
Recall the example:
\begin{align*}
F(x,y) &\longleftarrow G(x,z),H(y,z),I(x,y,z).
\end{align*}
We wish to maintain $F(x,y)$ given updated versions of the body
predicates $G',H',I'$.  We evaluate a maintenance rule of the form:
\begin{align*}
\delta F(x,y,\Delta) &\longleftarrow
\begin{array}[t]{l}
(\mathsf{Body}[G,H,I] \cdots \mathsf{Body}[G',H',I'])(x,y,z,\Delta), \\
\mathsf{ChangeOracle}(x,y,z).
\end{array}
\end{align*}
where $(\mathsf{Body}[G,H,I] \cdots \mathsf{Body}[G',H',I'])(x,y,z,\Delta)$
tabulates changes in satisfying assignments of the rule body,
and $\mathsf{ChangeOracle}(x,y,z)$ restricts evaluation 
to regions of the $(x,y,z)$ tuple space where changes may occur.
Roughly speaking, if you assert a new fact, the change-oracle
tells you where it \emph{could} be used; if you retract a fact, the
oracle tells you where it \emph{was} used.
The use of the change-oracle is crucial to efficiency.

The $\mathsf{ChangeOracle}$ predicate
is the essential heart of our maintenance algorithm.
During initial full-evaluation of the rule for $F$,
we build indices that note how changes to the predicates $G,H,I$
might affect evaluation.
To produce the $\mathsf{ChangeOracle}$ predicate, we
use the differences between the body predicates $(G\cdots G')$, $(H \cdots H')$
and $(I \cdots I')$ and these indices to produce the
$\mathsf{ChangeOracle}$ predicate.  Doing this efficiently
requires some special algorithms and data structures
described in \refsec{s:algorithms}.


Using delta-iterators (\refsec{s:deltaiterators}),
we can efficiently enumerate the changes
to the body predicates; let:
\begin{align*}
\delta G(x,z,\Delta) &= (G \cdots G')(x,z,\Delta) \\
\delta H(y,z,\Delta) &= (H \cdots H')(y,z,\Delta) \\
\delta I(x,y,z,\Delta) &= (I \cdots I')(x,y,z,\Delta)
\end{align*}
To construct the change-oracle, we need to determine what
portions of the $(x,y,z)$ tuple-space might need to be
revisited, given the changes $\delta G$, $\delta H$, and $\delta I$.
For this we use \emph{sensitivity indices} that record
how the rule evaluation is sensitive to changes in the body predicates.

\subsection{Sensitivity indices}

\label{s:sensitivityindices}

In our Datalog system, queries are evaluated using the leapfrog
triejoin algorithm (LFTJ) \cite{Veldhuizen:LB:2012}.
Here is an example leapfrog join for $A(x),B(x)$,
where $A = \{ 0,2,4,5,6\}$ and $B=\{1,2,6,7\}$:
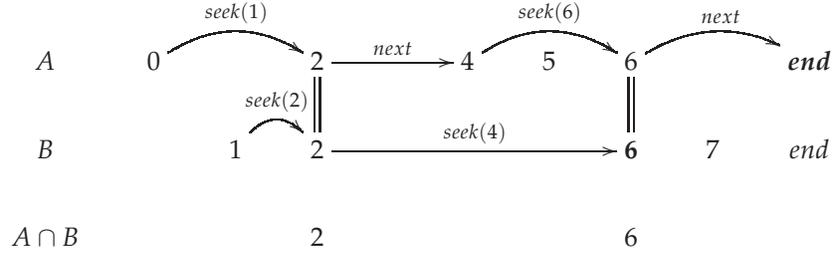
\begin{align*}
\xymatrix @=0.7cm {
A & 0 \ar@/^1pc/[rr]^{seek(1)} & & 2 \ar@{=}[d] \ar[rr]^{next} & & 4 \ar@/^1pc/[rr]^{seek(6)} & 5 & 6 \ar@{=}[d] \ar@/^1pc/[rr]^{next} & & \boldsymbol e \boldsymbol n \boldsymbol d \\
B & & 1 \ar@/^1pc/[r]^{seek(2)} & 2 \ar[rrrr]^{seek(4)} & & & & \boldsymbol 6 & 7 & end \\
A \cap B & & & 2 & & & & 6
}
\end{align*}
The leapfrog join begins by positioning
an iterator at the start of each predicate, then repeatedly
applying these rules (demonstrated by the arrows in the above diagram):
\begin{itemize}
\item If either iterator is at end, then stop.
\item If both iterators are positioned at the same key, emit this key.
Then increment one iterator.
\item Otherwise, take the iterator positioned at the lesser key, and
do a seek-lub to the key at which the other iterator is positioned.
\end{itemize}

We count the trace as the operations performed on the iterators, and their
result (e.g. one step in the above would be `seek(6) from x=4 to x=6 on iterator A').

LFTJ can handle most $\exists_1$ queries, but at the lowest level they
are implemented in terms of trie-iterator operations such as 
$\mathit{open()}$, $\mathit{up()}$, $\mathit{next()}$, and
$\mathit{seek\_lub()}$.
So, the approach we are about to describe for maintaining
$A(x),B(x)$ extends naturally to more complex queries.

We want to know: what changes to $A,B$ might cause changes to the trace?
For example, if we inserted $B(5)$, this would change the trace, since 
the seek(4) arrow from $x=2$ to $x=6$ in B would change to land on $x=5$.
However, if we inserted $B(3)$, this would not change the
trace, because the seek(4) arrow is seeking a least upper bound for 4;
the trace is not \emph{sensitive} to changes in B at $x=3$.


The rules for trace sensitivity of a unary predicate $D(x)$ are straightforward:
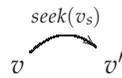
\begin{enumerate}
\item Seeks:
\begin{align*}
\xymatrix{
v \ar@/^1pc/[r]^{seek(v_s)} & v'
}
\end{align*}
If the iterator for predicate D is positioned at key $v$,
and a $\mathit{seek\_lub}(v_s)$ operation is performed so the iterator is then
positioned at $v'$, then the trace is sensitive to changes in
D in the interval $[v_s,v']$.  (It is not sensitive to changes
in $(v,v_s)$, because the seek operation finds a least upper
bound for $v_s$.)
\item Increment:
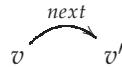
\begin{align*}
\xymatrix{
v \ar@/^1pc/[r]^{next} & v'
}
\end{align*}
If the iterator for D is positioned at key $v$,
and an increment (next) is performed so the iterator is then
positioned at $v'$, then the trace is sensitive to changes to
D in the interval $[v,v']$.
\item If the iterator for predicate $D$ is opened at position $v$
(i.e. the first record is $v$),
then the trace is sensitive to changes in $D$ in the interval
$(-\infty,v]$.
\end{enumerate}

For the above $A(x),B(x)$ example, the sensitivities are:
\begin{align*}
A_{sens} &= \{[-\infty,0], [1,2], [2,4], [6,6], [6,+\infty]\} \\
B_{sens} &= \{[-\infty,1], [2,2], [4,6] \}
\end{align*}
Given changes $\delta A$, $\delta B$, we collect intervals where
the predicate is sensitive \emph{and} a change has occurred there:
\begin{align*}
A_{co}([x_1,x_2]) &\longleftarrow \delta A(x,\Delta), x \in [x_1,x_2], A_{sens}([x_1,x_2]). \\
B_{co}([x_1,x_2]) &\longleftarrow \delta B(x,\Delta), x \in [x_1,x_2], B_{sens}([x_1,x_2]).
\end{align*}
To evaluate the above rules for $A_{co},B_{co}$ efficiently, we use
an interval-tree representation for $A_{sens}$ and 
$B_{sens}$ (\refsec{s:intervaltrees}).

We can then define the change-oracle as:
\begin{align*}
\mathsf{ChangeOracle}(x) &\equiv
x \in [x_1,x_2], \left( A_{co}([x_1,x_2]); B_{co}([x_1,x_2]) \right).
\end{align*}
(In our notation, the semicolon indicates a disjunction.)
For efficiency, we treat $\mathsf{ChangeOracle}$ as a
nonmaterialized predicate: during evaluation of the
maintenance rule, the iterator for
$\mathsf{ChangeOracle}(x)$ is internally
manipulating the $A_{co}$ and $B_{co}$ predicates to present
the contents of $\mathsf{ChangeOracle}(x)$, without
explicitly expanding the intervals into individual
elements.

A note on the maintenance cycle:
after each matching interval
in $A_{sens}$ is found,
we remove it from $A_{sens}$; this guarantees that the cost
of evaluating the $A_{co}$ rule is $O\left((|\delta A| + |A_{co}|) \log n\right)$,
i.e. proportional to the number of changes and $A_{co}$-results.
The $\log n$ factor reflects the btree heights; a sharper estimate
would be to take $n = \max(|\delta A|, |A_{sens}|)$.
When the maintenance rule is evaluated, we accumulate new sensitivity
intervals to $A_{sens}$ and $B_{sens}$, so we are ready for
the next round of maintenance.

\subsubsection{Example of maintenance for a unary join}

For a concrete example, suppose we have:
\begin{align*}
\delta A &= \{ (5,\mathrm{ERASE}), (8,\mathrm{INSERT}) \} \\
\delta B &= \{ (2,\mathrm{ERASE}), (3,\mathrm{INSERT}) \}
\end{align*}


This diagram shows the changes made to A and B, and the sensitivity
intervals:
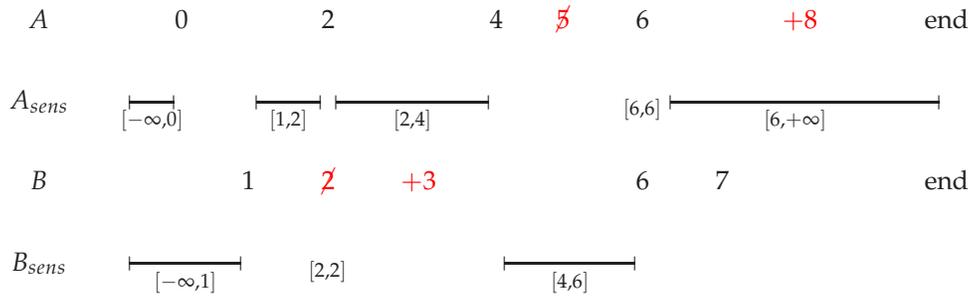
\begin{align*}
\xymatrix @=0.5cm {
A & & 0 & & 2 & & 4 & {\color{red}\cancel{5}} & 6 & & {\color{red}+8} & & \mathrm{end} \\
A_{sens}  & \ar@{|-|}[r]_{[-\infty,0]} & & \ar@{|-|}[r]_{[1,2]} & \ar@{|-|}[rr]_{[2,4]} & & & & {}_{[6,6]} \ar@{|-|}[rrrr]_{[6,+\infty]} & & & & \\
B & & & 1 & {\color{red}\cancel{2}} & {\color{red}+3} & & & 6 & 7 & & & \mathrm{end} \\
B_{sens} & \ar@{|-|}[rr]_{[-\infty,1]} & & & {}_{[2,2]} & & \ar@{|-|}[rr]_{[4,6]} & &  & & & 
}
\end{align*}

When we evaluate the rules for $A_{co},B_{co}$, we find these
contributions:
\begin{align*}
\begin{array}{lc}
\mathrm{change} & \mathrm{contributions~~to~~}A_{co},B_{co} \\
A: (5,\mathrm{ERASE}) & \emptyset \\
A: (8,\mathrm{INSERT}) & \{ [6,\mathsf{end}] \} \\
B: (2,\mathrm{ERASE}) & \{ [2,2] \} \\
B: (3,\mathrm{INSERT}) & \emptyset
\end{array}
\end{align*}

and so
\begin{align*}
\mathsf{ChangeOracle}(x) &= \bigcup \left\{ [2,2], [6,+\infty] \right\}
\end{align*}

Maintenance: because of the ChangeOracle, we skip immediately to $x=2$;
there we find that $2$ is no longer in $A \cap B$.  Then we skip to the
start of the next interval in the ChangeOracle, $x \in [6,+\infty]$:
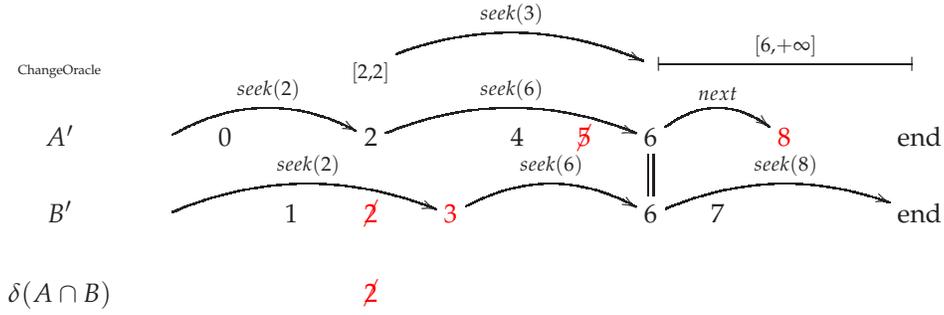
\begin{align*}
\xymatrix @=0.5cm {
\txt{\tiny ChangeOracle} & & & & {}_{[2,2]} \ar@/^1pc/[rrrr]^{seek(3)} & & & & \ar@{|-|}[rrrr]^{[6,+\infty]} & & & & \\
A' & \ar@/^1pc/[rrr]^{seek(2)} & 0 & & 2 \ar@/^1pc/[rrrr]^{seek(6)} & & 4 & {\color{red}\cancel{5}} & 6 \ar@/^1pc/[rr]^{next} \ar@{=}[d] & & {\color{red}8} & & \mathrm{end} \\
B' & \ar@/^1pc/[rrrr]^{seek(2)} & & 1 & {\color{red}\cancel{2}} & {\color{red}3} \ar@/^1pc/[rrr]^{seek(6)} & & & 6 \ar@/^1pc/[rrrr]^{seek(8)} & 7 & & & \mathrm{end} \\
\delta (A \cap B) & & & & {\color{red}\cancel{2}} & & & & & & 
}
\end{align*}

During evaluation of the maintenance rule, the sensitivity intervals
are updated, so we are ready for the next round of maintenance:
the intervals we examined because of the ChangeOracle are removed, and we
insert new ones due to iterator operations as we evaluate the rule.
The revised sensitivity intervals are:
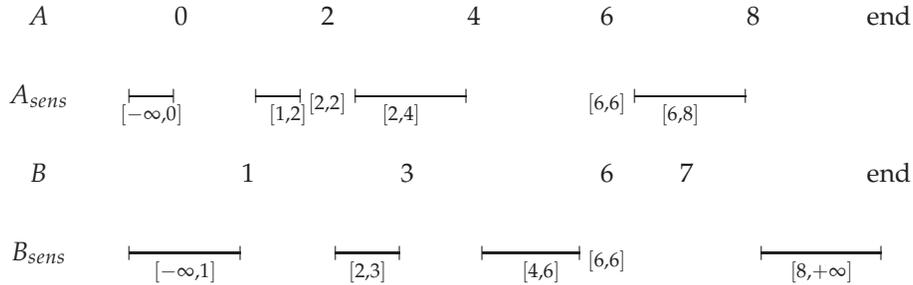
\begin{align*}
\xymatrix @=0.5cm {
A & & 0 & & 2 & & 4 &  & 6 & & 8 & & \mathrm{end} \\
A_{sens}  & \ar@{|-|}[r]_{[-\infty,0]} & & \ar@{|-|}[r]_{[1,2]} & {}_{[2,2]} \ar@{|-|}[rr]_{[2,4]} & & & & {}_{[6,6]} \ar@{|-|}[rr]_{[6,8]} & & & & \\
B & & & 1 &  & 3 & & & 6 & 7 & & & \mathrm{end} \\
B_{sens} & \ar@{|-|}[rr]_{[-\infty,1]} & & & \ar@{|-|}[r]_{[2,3]} & & \ar@{|-|}[rr]_{[4,6]} & & {}_{[6,6]} & & \ar@{|-|}[rr]_{[8,+\infty]} & &
}
\end{align*}

\subsection{Sensitivity indices for predicates with multiple arguments}

For a trivial query like $A(x),B(x)$, sensitivity indices \& the 
change-oracle are of marginal use; in fact our system would not use
them for such a simple query.  However, for complex queries these
techniques make a tremendous difference.

Consider this example:
\begin{align*}
F(x,y) &\longleftarrow G(x,z),H(y,z),I(x,y,z),R(z).
\end{align*}

Suppose the optimizer chooses the key-variable ordering $[x,y,z]$.
If a fact is retracted from $R(z)$, the change-oracle lets us
examine only those $(x,y,z)$ tuples where that fact was used.

The approach to building the change-oracle for predicates
with multiple key arguments generalizes that described
in the previous section.
First, a bit of background.

Recall that LFTJ evaluates rules using a `backtracking search through
tuple space,' which conceptually consists of nested
leapfrog joins on unary predicates.  We write $G(x,\_)$ for projection,
and $G_x(y)$ for a curried version of $G$ for a specific
$x$.  For instance, given $G=\{ (0,10), (0,20), (1,30) \}$,
we would have:
\begin{align*}
G(x,\_) &= \{ 0, 1 \} \\
G_0(y) &= \{ 10, 20 \} \\
G_1(y) &= \{ 30 \}
\end{align*}
(Note: we do not explicitly construct these projections
and curried versions; this is just for exposition.)

The LFTJ algorithm does a backtracking search through 
the $[x,y,z]$ space, first seeking a binding for $x$, then
proceeding to a binding for $y$ once $x$ is found, etc.
Conceptually, the three nested queries used are:
\begin{enumerate}
\item $G(x,\_),I(x,\_,\_)$
\item $H(y,\_),I_x(y,\_)$
\item $G_x(z),H_y(z),I_{xy}(z),R(z)$
\end{enumerate}

When we evaluate the query, we record sensitivity information
much as described earlier for unary predicates.  However,
we also record information about the bindings of other
key-variables.  The sensitivity predicate for $R$, for
instance would have the form
$R_{sens,z}([z_1,z_2],x,y)$.  If a fact is removed from $R$,
we can quickly determine the $(x,y,z)$ bindings where that
fact was used; if a fact is added, we can quickly determine
where it could be used.

The sensitivity predicate for $H_y(z)$ illustrates the general form:
\begin{align*}
H_{sens,z}(\underbrace{y}_{(1)},\underbrace{[z_1,z_2]}_{(2)},\underbrace{x}_{(3)})
\end{align*}
In position (1) we have variables that precede $z$ in the argument
list for $H$; in position (2) we have the sensitivity interval for $z$;
in position (3) we have key-variables that are bound before $z$ but do
not appear in the argument-list for $H$.

So, \emph{conceptually}, we would have these sensitivity predicates:
\begin{align*}
\begin{array}{l}
G_{sens,x}([x_1,x_2]) \\
G_{sens,z}(x,[z_1,z_2],y) \\
H_{sens,y}([y_1,y_2],x) \\
H_{sens,z}(y,[z_1,z_2],x) \\
I_{sens,x}([x_1,x_2]) \\
I_{sens,y}(x,[y_1,y_2]) \\
I_{sens,z}(x,y,[z_1,z_2]) \\
R_{sens,z}([z_1,z_2],x,y)
\end{array}
\end{align*}

In practice we can drop sensitivity indices where the key-arguments of
the atom form a prefix of the key-ordering chosen by the optimizer.  For
example, our implementation would not bother creating sensitivity indices
for $I$, since its arguments match the chosen key order $[x,y,z]$; it would also
not create the index $G_{sens,x}([x_1,x_2])$, since $(x)$ is also
a prefix of $[x,y,z]$.

\subsubsection{Tree surgery operations}

The delta-iterator described in \refsec{s:deltaiterators} lets us
efficiently enumerate the changed records between two consecutive
versions of a predicate.  For building the change oracle, we need
finer information, namely, changes made to the \emph{trie} presentation
of the predicate.  We call such changes \emph{tree surgery operations}.
Tree surgery operations consist of either inserting or removing branches.

For example, consider these two versions of a predicate $A(x,y,z)$:

\begin{centering}
\begin{tabular}{c|c}
Version 1  & Version 2 \\ \hline
(0,30,80)  &  (0,30,80) \\
(0,30,81)  & \\
(1,35,60)  &  (1,35,60) \\
(1,35,61)  &  (1,35,61) \\
(3,40,90)  & \\
(3,50,91)  &  (3,50,91) \\
(3,50,92)  & \\
           &  (4,60,71)
\end{tabular}

\end{centering}

The delta-iterator would produce this stream of changes:

\begin{centering}
\begin{tabular}{ll}
ERASE & 0,30,81 \\
ERASE & 3,40,90 \\
ERASE & 3,50,92 \\
INSERT & 4,60,71
\end{tabular}

\end{centering}

The trie presentations of the two versions are:

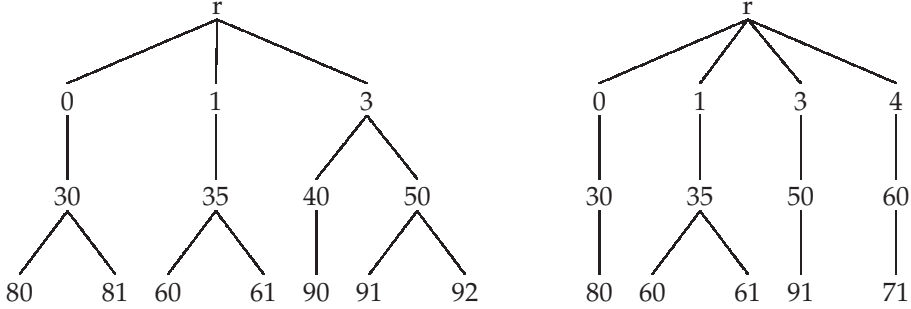
\begin{minipage}{3in}
\synttree[r[0[30[80][81]]][1[35[60][61]]][3[40[90]][50[91][92]]]]
\end{minipage}
\begin{minipage}{3in}
\synttree[r[0[30[80]]][1[35[60][61]]][3[50[91]]][4[60[71]]]]
\end{minipage}

The tree surgery operations would be:

\begin{centering}
\begin{tabular}{ll}
ERASE & 0--30--81 \\
ERASE & 3--40--90 \\
ERASE & 3--40 \\
ERASE & 3--50--92 \\
INSERT & 4 \\
INSERT & 4--60 \\
INSERT & 4--60--71
\end{tabular}

\end{centering}

It is reasonably straightforward and efficient to adapt a
delta-iterator into an iterator of tree-surgery operations,
with a little bookkeeping; we omit the details here.

\subsubsection{Matching tree surgery operations with sensitivity indices}

Returning to our running example, recall that we have these two sensitivity
indices for $H(y,z)$:
\begin{align*}
H_{sens,y}([y_1,y_2],x) \\
H_{sens,z}(y,[z_1,z_2],x)
\end{align*}
We use a tree-surgery adaptor to get the changes made to the trie
presentation of $H$, from the delta-iterator giving us the changes
in $H$.  Tree surgeries on $H$ come in two forms: those that insert
or remove vertices at depth 1, and those that insert or remove vertices
at depth 2.  We collect these surgeries by depth, writing
$\delta H^1(y,\Delta)$ and $\delta H^2(y,z,\Delta)$ for
depth-1 and depth-2 surgeries, respectively.

Trie surgery operations of depth 1 are matched with
intervals in $H_{sens,y}([y_1,y_2],x)$, and trie surgery operations
of depth2 are matched to intervals in $H_{sens,z}(y,[z_1,z_2],x)$.
The resulting change-oracle contributions we call $H_{co,y}$
and $H_{co,z}$, and are defined by:
\begin{align*}
H_{co,y}(x,[y_1,y_2]) &\longleftarrow \delta H^1(y,\Delta),y \in [y_1,y_2],H_{sens,y}([y_1,y_2],x) \\
H_{co,z}(x,y,[z_1,z_2]) &\longleftarrow \delta H^2(y,z,\Delta),z \in [z_1,z_2],H_{sens,z}([y,[z_1,z_2],x)
\end{align*}
As mentioned previously, this is implemented with interval trees and
is very efficient---proportional (modulo $\log n$) to the number of
tree-surgery operations plus the number of matches to those operations
in the sensitivity indices (\refsec{s:intervaltrees}).
(Also recall that we remove matched intervals from the sensitivity indices.)

We define contributions to the change oracle from $G$, $I$, and $R$ similarly.
Finally, we define the change oracle by these (nonmaterialized) definitions:
\begin{align*}
\mathsf{ChangeOracle}(x,y,z) &\longleftarrow \mathsf{ChangeOracle}_1(x); \mathsf{ChangeOracle}_2(x,y); \mathsf{ChangeOracle}_3(x,y,z). \\
\mathsf{ChangeOracle}_1(x) &\longleftarrow (G_{co,x}([x_1,x_2]); I_{co,x}([x_1,x_2])), x \in [x_1,x_2]. \\
\mathsf{ChangeOracle}_2(x,y) &\longleftarrow (H_{co,y}(x,[y_1,y_2]); I_{co,y}(x,[y_1,y_2])), y \in [y_1,y_2]. \\
\mathsf{ChangeOracle}_3(x,y,z) &\longleftarrow
   (
\begin{array}[t]{l}
G_{co,z}(x,y,[z_1,z_2]); H_{co,z}(x,y,[z_1,z_2]); \\
I_{co,z}(x,y,[z_1,z_2]); R_{co,z}(x,y,[z_1,z_2])), z \in [z_1,z_2].
\end{array}
\end{align*}

We can then maintain the rule, using the maintenance rule:
\begin{align*}
\delta F(x,y,\Delta) &\longleftarrow
\begin{array}[t]{l}
(\mathsf{Body}[G,H,I] \cdots \mathsf{Body}[G',H',I'])(x,y,z,\Delta), \\
\mathsf{ChangeOracle}(x,y,z).
\end{array}
\end{align*}
Recall that when evaluating the maintenance rule, we accumulate new intervals
to the sensitivity indices, so we are ready for the next round of maintenance.

\bibliography{bibliography}
\bibliographystyle{plain}

\end{document}